\documentclass[traditabstract]{aa}

\usepackage{graphicx}
\usepackage{amsmath}
\usepackage{natbib}
\usepackage{ulem}

\newcommand{\uloop}{$U$ loop}
\newcommand{\iuloop}{inverse-$U$ loop}
\newcommand{\uu}{$U$}
\newcommand{\iuu}{inverse-$U$}

\begin{document}
\title{Decay of a simulated mixed-polarity magnetic field 
       in the solar surface layers}
\author{R. Cameron\inst{1}
\and A. V\"ogler\inst{2}
\and M. Sch\"ussler\inst{1}
}
\titlerunning{Decay of a simulated mixed-polarity field}
\institute{Max-Planck-Institut f\"ur Sonnensystemforschung, Katlenburg-Lindau
\and Sterrenkundig Instituut, Utrecht University, Postbus 80000, 3508 TA Utrecht, The Netherlands}
\date{Received; accepted}
\abstract{Magnetic flux is continuously being removed and replenished on the solar 
surface. To understand the removal process we carried out  3D radiative MHD simulations of the
evolution of patches of photospheric magnetic field with equal amounts of positive 
and negative flux. 
We find that the flux is removed at a rate corresponding to an effective turbulent
diffusivity, $\eta_{\mathrm{eff}}$, of 100--340 km$^2$s$^{-1}$, depending on the boundary conditions. 
For average unsigned flux densities
above about 70 Gauss, the percentage of surface magnetic energy coming from different
field strengths is almost invariant. The overall process is then one where magnetic 
elements are advected by the horizontal granular motions and occasionally come into contact with 
opposite-polarity elements. These reconnect above the photosphere on a comparatively short 
time scale after which the {\uloop}s produced rapidly escape through the upper surface while
the downward retraction of {\iuloop}s is significantly slower, because of 
the higher inertia and lower plasma beta in the deeper layers.}
\maketitle
\section{Introduction}
High-resolution spectropolarimetric observations \citep[eg][]{De_Pontieu02, Martinez_Gonzalez07,
Centeno07, Ishikawa10, Kubo10, Solanki10, Borrero10, Danilovic10} 
show that magnetic flux on the solar surface is continuously 
removed and replenished. The small scale fields involved are probably a mixture of 
flux which is left over from the decay of active regions and flux generated
in the near-photospheric layers by local-dynamo action \citep{Hagenaar03,Voegler07, Pietarila_Graham09,
 Pietarila_Graham10}. 
This field has been described as 'salt and pepper', reflecting small spatial scales on which 
the polarities are mixed.
Reconnection events take 
place when the surface magnetic fields of opposite polarities, which are continually 
advected by the granular motions, come into close proximity. If we consider only a 
layer near the surface, then a field line in the layer can be classified as a {\uloop}
if both ends of the field line pass through the upper boundary of the layer, and as a {\iuloop}
if both ends pass through the lower boundary. Reconnection of field lines initially
passing from the bottom of the layer to the top with field lines going from the top
to the bottom produces pairs of {\uu} and {\iuu} loops. Magnetic tension then causes 
the loops to retract: {\iuloop}s downwards through the photosphere  and {\uloop}s 
upwards. Observations indicate that the reconnection occurs above the optical surface, so that the 
removal of flux from the surface involves the subsequent retraction of
{\uloop}s \citep{Kubo10}.

In this paper we study the evolution of a small-scale mixed polarity field using numerical
simulations. We concentrate not so much on individual reconnection events but on the
statistical properties of the flux removal such as how fast 
does the removal occur, what are the processes which determine the rate, what are the 
dynamics of {\uu} and {\iuu} loops. The observational signature of individual reconnection 
events has been studied by \cite{Danilovic09}.

This paper is organized as follows. In Section 2 we briefly sketch the
numerical code we have used and the simulation setup. Section 3 presents 
the simulation results and their dependence on the assumed initial and boundary conditions.
Section 4 discusses the results, and Section 5 gives our conclusions.

\section{Simulations}
Our simulations consider the near-surface layers of the Sun 
extending  from about 800~km below the optical surface
to 600~km above. The physics is described by the continuity and momentum
equations, the induction equation, an energy equation which includes radiative
transfer, and an equation of state which allows for the effects of partial
ionization. The equations are fully set out in \cite{Voegler05},
where also the details of the MURaM code we have used for this study are described.

The code has been used to investigate numerous aspects of solar-surface
magnetic activity, including quiet-Sun magnetism 
\citep{Keller04,Khomenko05,Voegler07,Pietarila_Graham09, Pietarila_Graham10, Danilovic10a, Danilovic10b}
pores \citep{Cameron07}, emerging flux \citep{Cheung08, Yelle_Chaouche09, Cheung10}, 
and sunspots \citep{Schuessler06, Rempel09, Rempel09b}. 

For this study we first consider a reference case with a box size of
6~Mm$\times$6~Mm in the $x$ and $y$  (horizontal) directions 
and 1.4~Mm in the $z$ (vertical) direction with a grid resolution of 20.8~km in each of 
the horizontal directions and 14~km in the vertical. The bottom boundary
condition allows for in- and outflows as described in \cite{Voegler05}. The magnetic field is vertical
at the top boundary, where the vertical component of the velocity vanishes and a free-slip condition
(vertical derivatives vanish) is condition is imposed on the horizontal velocity components.
The box is periodic in both horizontal directions.
The initial condition for our reference case is a height-independent checkerboard ($2\times2$) vertical magnetic
field, i.e.
\begin{equation}
B_z=\begin{cases}
   -200\textrm{G}&\text{for $0<x<3$~Mm and $0<y<3$~Mm}\\
   -200\textrm{G}&\text{for $3<x<6$~Mm and $3<y<6$~Mm}\\
   +200\textrm{G}&\text{for $3<x<6$~Mm $0<y<3$~Mm}\\
   +200\textrm{G}&\text{for $0<x<3$~Mm $3<y<6$~Mm}.
\end{cases}\\
\end{equation}
\begin{equation*}
B_x=B_y=0,
\end{equation*}
superposed on a well developed non-magnetic convection simulation.
This initial condition is not particularly realistic in that we do not expect 
purely vertical fields to occur over a region this large in the quiet Sun. The initial condition
is instead designed to investigate the manner in which flux in the quiet Sun is dispersed,
reconnects and is removed from the surface.

We also consider several variations of this setup. We study the effect of the
initial condition by starting from a magnetic field configuration consisting of
two stripes ($2\times 1$) of opposite polarity: 
\begin{equation}
B_z=\begin{cases}
   -200G&\text{for $0<x<3$~Mm}\\
   +200G&\text{for $3<x<6$~Mm}.
\end{cases}
\end{equation}

We also investigate the sensitivity to the upper boundary condition by comparing the reference
case to ones where the magnetic field was matched to a potential field at the upper boundary,
to cases where the top boundary is raised by additional 280 km above the optical surface, and to cases
where the top boundary is open to flows. These closed and open upper boundary conditions either
completely or partially reflect outgoing waves. The influence of the reflected waves, and of waves otherwise 
excited in the chromosphere, on the retraction or escape of
the loops is not studied beyond using these two different boundary conditions.
Finally we consider the effect of varying the 
magnetic diffusivity, $\eta$.

\section{Results}
\subsection{2$\times$2 case}
\begin{figure}[h!]
 \centering
\resizebox{0.95\hsize}{!}{\includegraphics{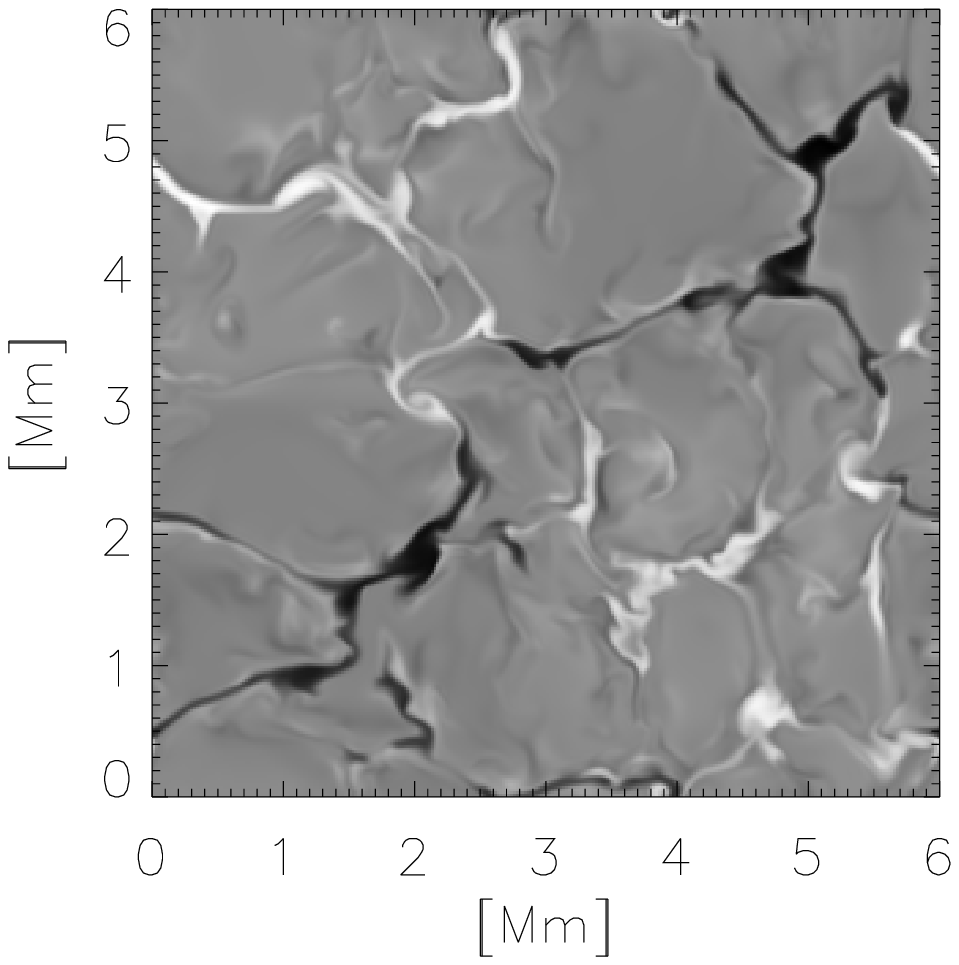}\includegraphics{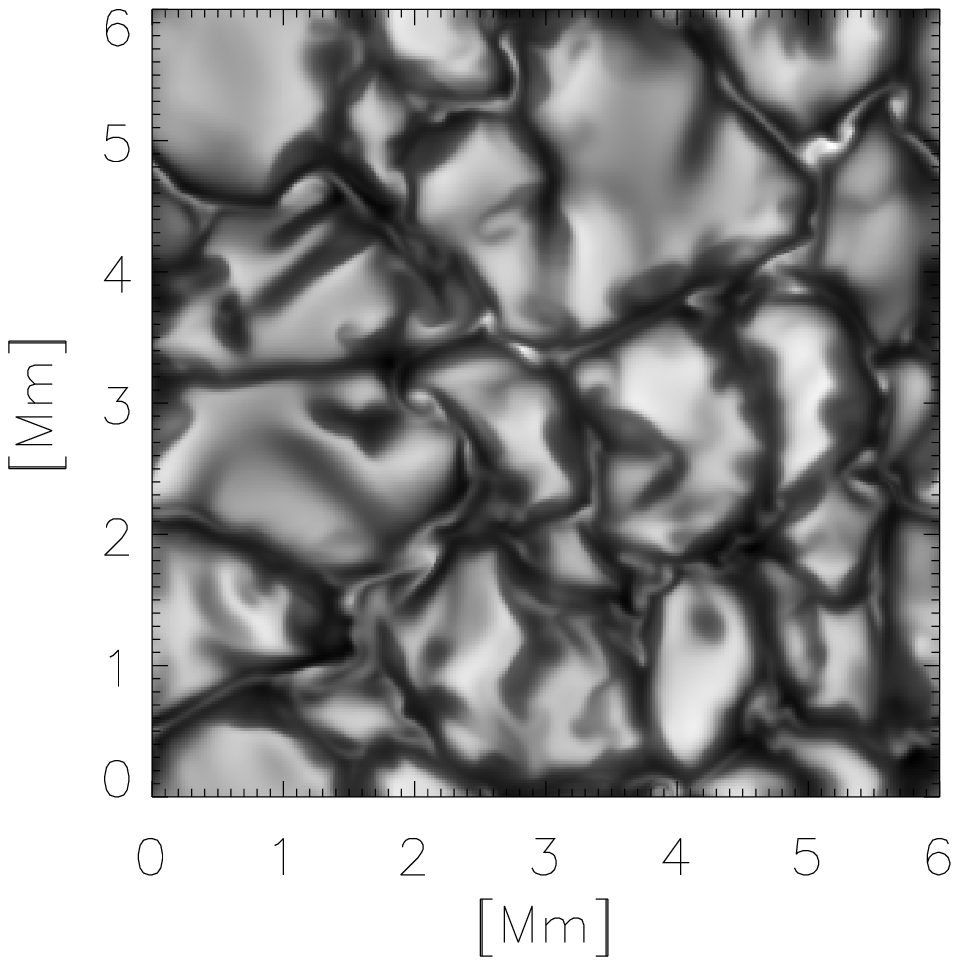}}
\resizebox{0.95\hsize}{!}{\includegraphics{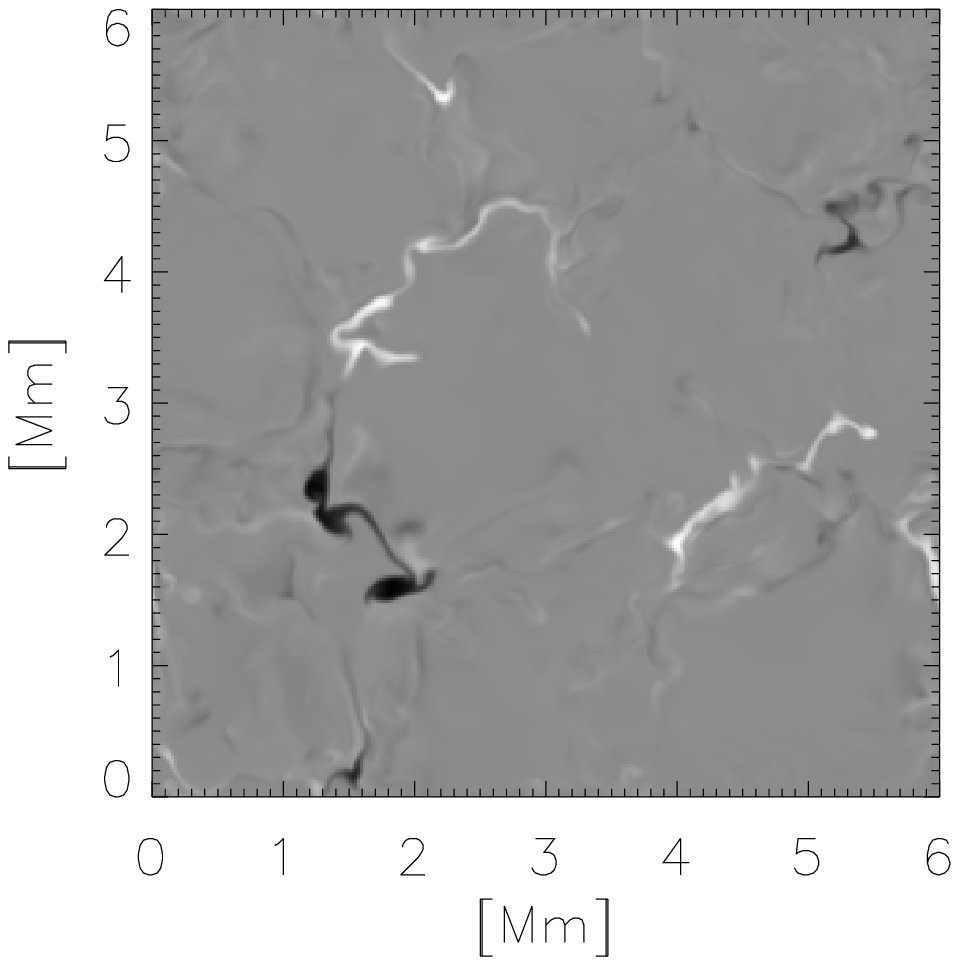}\includegraphics{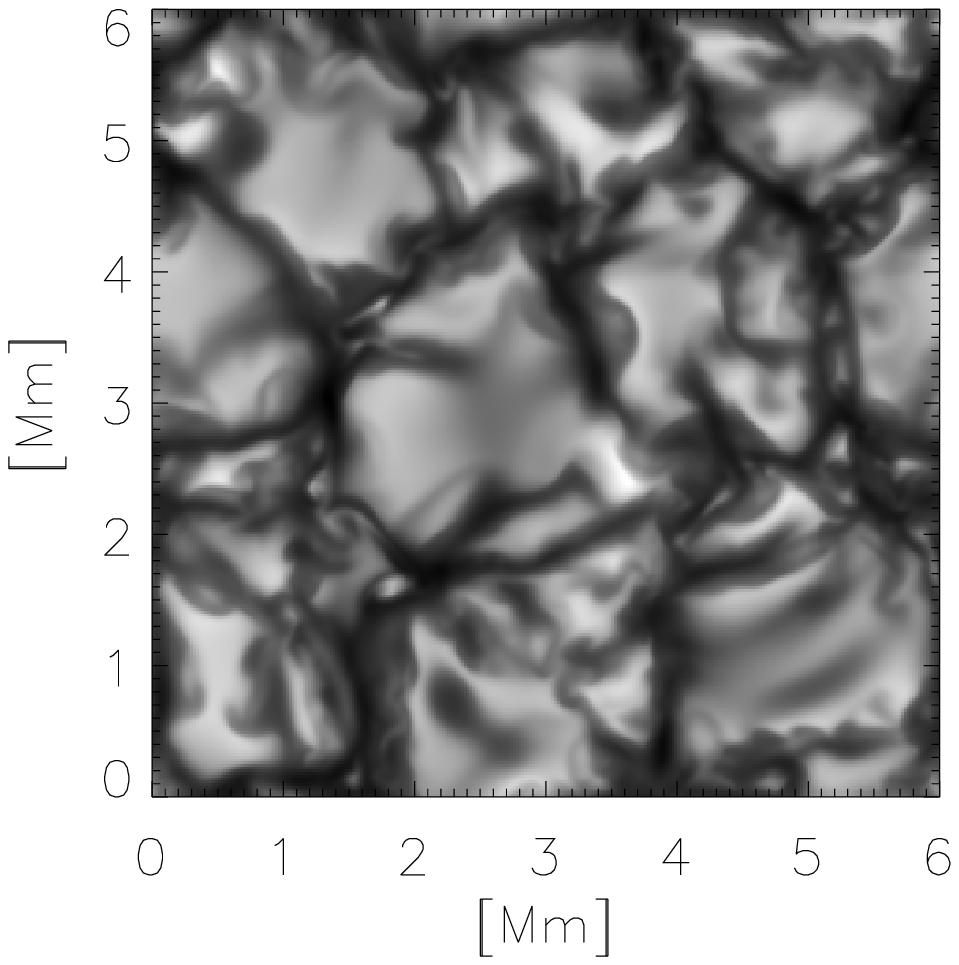}}
\resizebox{0.95\hsize}{!}{\includegraphics{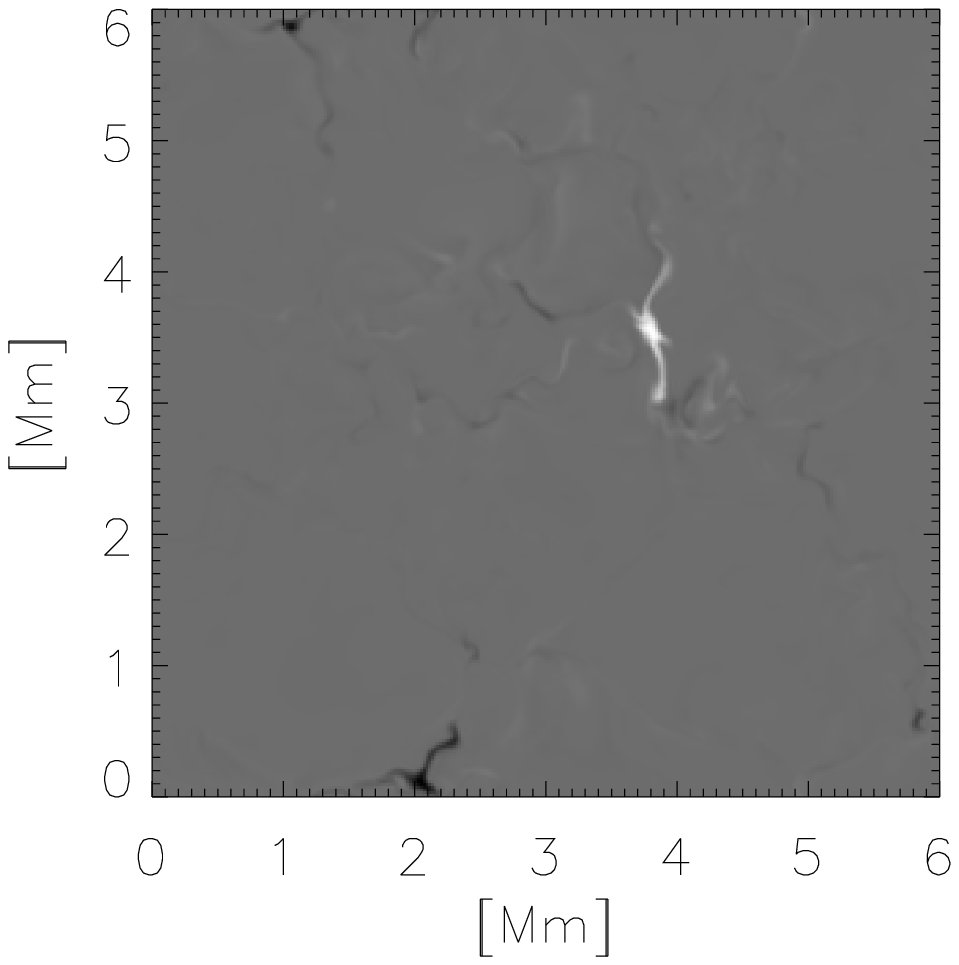}\includegraphics{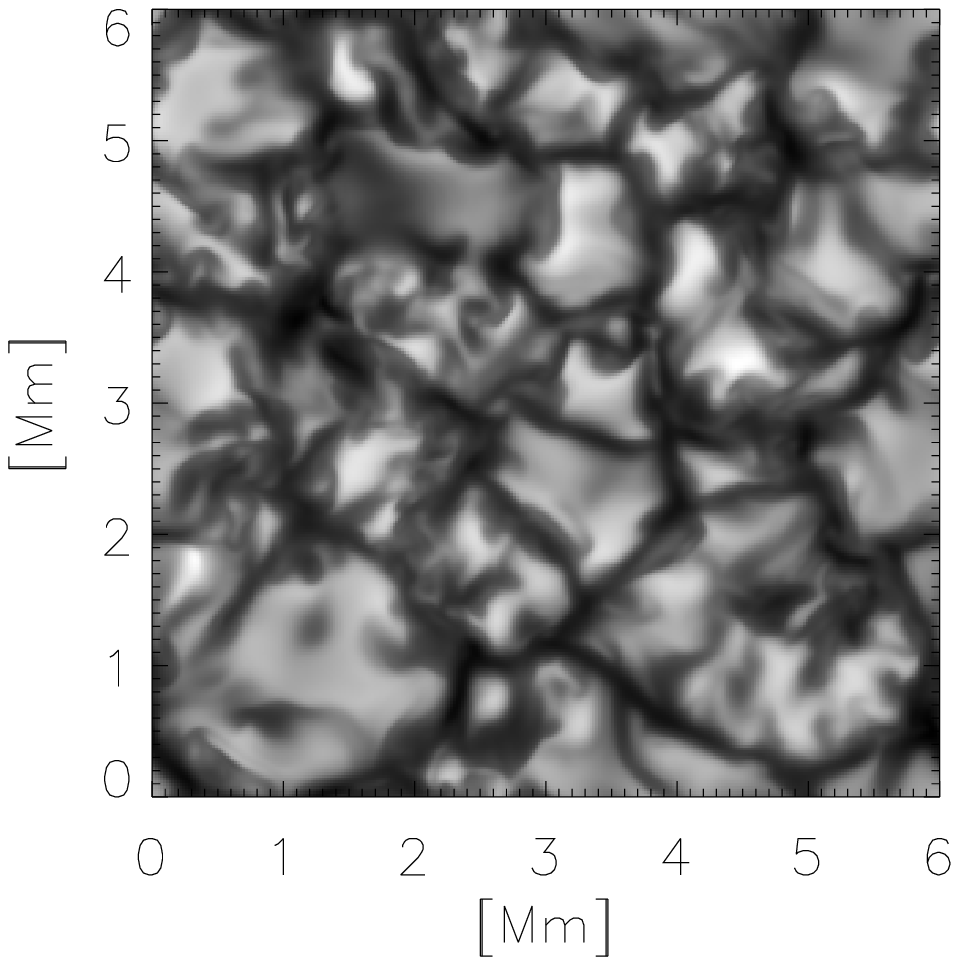}}
\caption{Time evolution of $B_z$ on the plane $z=0$, corresponding to the
average height of the optical surface (left panels) and the bolometric brightness
(right panels) for the reference ($2\times 2$) case
at 4 min (top), 80 min (middle), and 200 min (bottom) after introduction
of the initial field. The average unsigned magnetic field for these snapshots
corresponds to about 200~G, 50~G, and 10~G, respectively.
}
\label{fig:evolution-bz}
\end{figure}

We take as our reference case the simulation starting from  the checkerboard (2$\times$2) 
initial condition, with vertical magnetic field boundary condition at the upper boundary, 
horizontal resolution of $20.8$~km, and $\eta=11$~km$^2$s$^{-1}$. The magnetic Reynolds number 
of these simulations is  below the critical value required for small-scale dynamo action \citep{Voegler07}.
This means that the simulations are too diffusive to show local dynamo action.
Since our boundary conditions do not allow for incoming flux at the 
lower boundary, the magnetic field must inevitably decay. This can be
seen qualitatively in Fig.~\ref{fig:evolution-bz}, which shows maps of the vertical magnetic field
component on the horizontal surface $z=0$, at the average (time and space) 
height of the optical ($\tau=1$) surface. The temporal variation of the spatially averaged $\tau=1$ height     
for this particular run was calculated for four different times and was found to have 
rms deviations of 8.5~km. For comparison the spatial rms deviations at any one time are about 4 times larger
so that the $\tau=1$ surface can be regarded as a corrugated suface whose average height only
varies slightly.


\begin{figure}[h!]
\resizebox{1.0\hsize}{!}{ 
\includegraphics{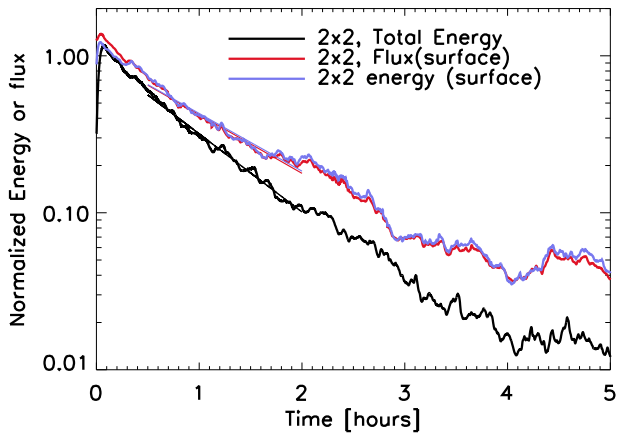}}
\caption{Decay of total magnetic energy in the entire simulation box,   
and of unsigned vertical magnetic flux at $z=0$ for the reference case.  
The thin lines are exponential fits. All
curves have been normalized such that the fits have a value of unity at $t=0$.}
\label{fig:Time_evol_ref}
\end{figure}

To quantitatively analyze the decay we consider the time evolution of the 
magnetic energy in a horizontal cut at a fixed geometrical height near $z=0$, 
the evolution of the unsigned vertical magnetic flux through the same surface, and 
the magnetic energy in the whole computational domain. We chose these three 
quantities because, 1) the vertical flux is the simplest quantity most relevant for
comparison with observations, 2) the total magnetic energy because it gives information 
on the global properties of the simulation,  and 3) the magnetic energy at the height of the 
vertical flux slice gives insight into the connection between previous
two. 
The evolution of these quantities is shown
in Figure~\ref{fig:Time_evol_ref}.

\begin{figure}[h!]
\resizebox{1.0\hsize}{!}{\includegraphics{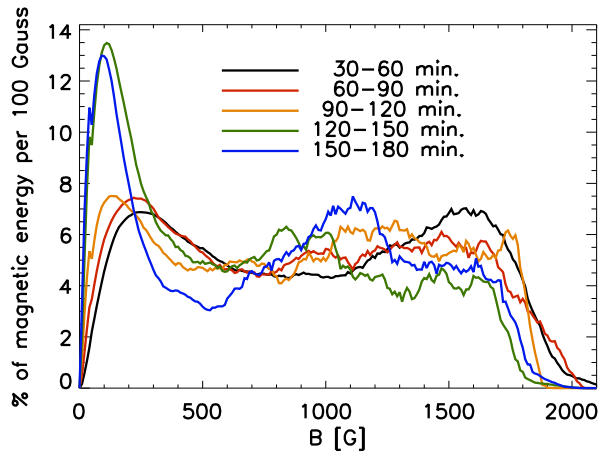}}
\caption{The percentage contribution to the total surface magnetic energy density per
100 Gauss, at different field strengths for different times (2$\times$2 reference case).}
\label{fig:energy-histogram}
\end{figure}

An interesting feature of Figure~\ref{fig:Time_evol_ref} is that
the unsigned flux, and the magnetic energy density at $z=0$ almost evolve in parallel. 
This runs against the naive expectation 
that if the magnetic field decays at a rate $\alpha$, so that 
$B\sim \exp(-\alpha t)$ then $B^2\sim \exp(-2\alpha t)$,
suggesting that the magnetic energy should decay twice as fast 
as the unsigned magnetic flux. The reason why this relation 
does not hold can be seen qualitatively in Figure~\ref{fig:evolution-bz}:
a significant amount of the magnetic flux is concentrated in a number of features
and it is the number of such features, rather than the field strength in the
features, which decreases in time. Therefore both the unsigned flux and the energy
are proportional to the number of features and thus decay in parallel.
This interpretation is confirmed in Figure~\ref{fig:energy-histogram},
where we see that the percentage of surface magnetic energy corresponding to the different
field strengths varies only weakly for the first 2 hours. In the third
hour we begin to see an increasing contribution from weak field regions, as
the number of strong-field elements diminishes.  
In particular we note that at 200 minutes
there are only two reasonably strong magnetic features, one of each polarity, 
left (cf. Fig.~\ref{fig:evolution-bz}, bottom panels).

Returning to Figure~2, the total magnetic energy in the entire computational
domain decays faster than the surface energy densities and flux.
We have estimated the decay rates by fitting an exponential function over the 
period from $t=30$~min to $t=120$~min. This period
was chosen so that the initial phase, during which flux expulsion acts to 
increase the energy, is excluded. 
The $e$-folding time for the surface flux and magnetic energy,
derived from the fits, is 70 minutes. 
The $e$-folding time for the magnetic energy over the entire domain is
52 minutes. The ratio between the decay rates of surface and volume integrated
quantities is thus approximately 0.74, which is intermediate between 
the naive expectation of 0.5 and the expectation from magnetic
elements of 1. This indicates that the magnetic structures
become more volume filling (less intermittent) with depth.

To characterize the decay in terms of a turbulent diffusivity, 
we consider the two-dimensional linear diffusion equation, 
${\partial B_z}/{\partial t}=\eta_{\mathrm {eff}} \nabla^2 B_z$, 
where $B_z$ is the vertical component of the field on the plane $z=0$.
An assumption here is that the vertical flux diffuses as a passive scalar.
The slowest decaying eigenmode with the symmetries of the initial condition
has a spatial dependence  $\sin(2\pi x/L) \sin(2\pi y/L)$ where 
$L=6000/\sqrt{2}$~km is the wavelength. The $e$-folding rates determined from the simulation
can then be used to 
evaluate $\eta_{\mathrm {eff}}= 145$~km$^2$s$^{-1}$, which is about a factor of 
13 larger than the explicit diffusivity in the simulation.

The reconnection of the magnetic structures takes place when 
magnetic structures of opposite polarity come into close proximity.

\begin{figure*}[!]
\centering
\includegraphics[width=\hsize]{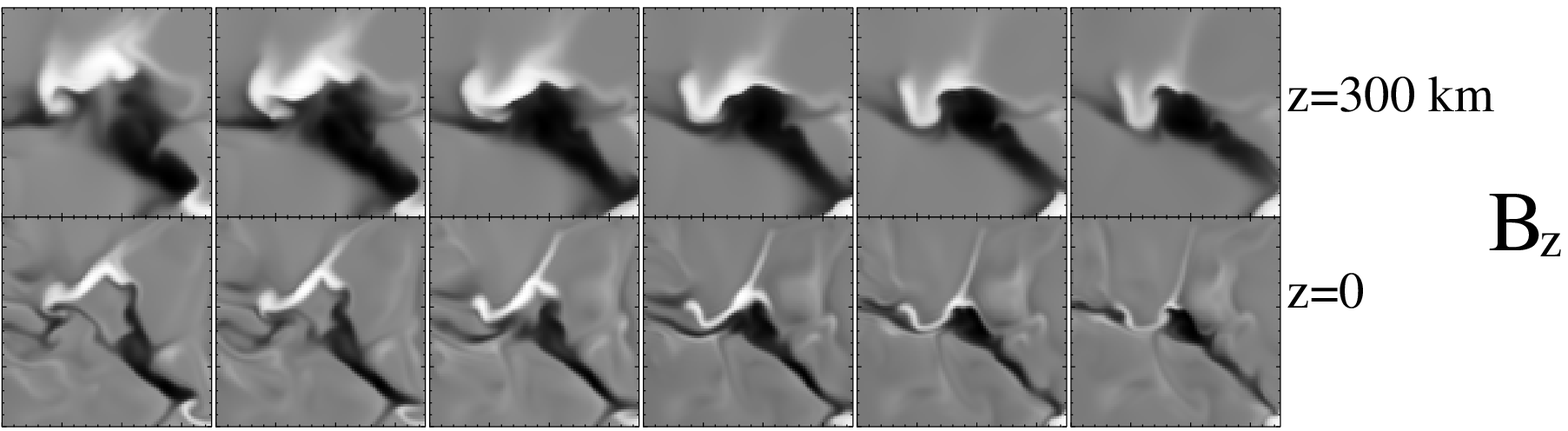}
\includegraphics[width=\hsize]{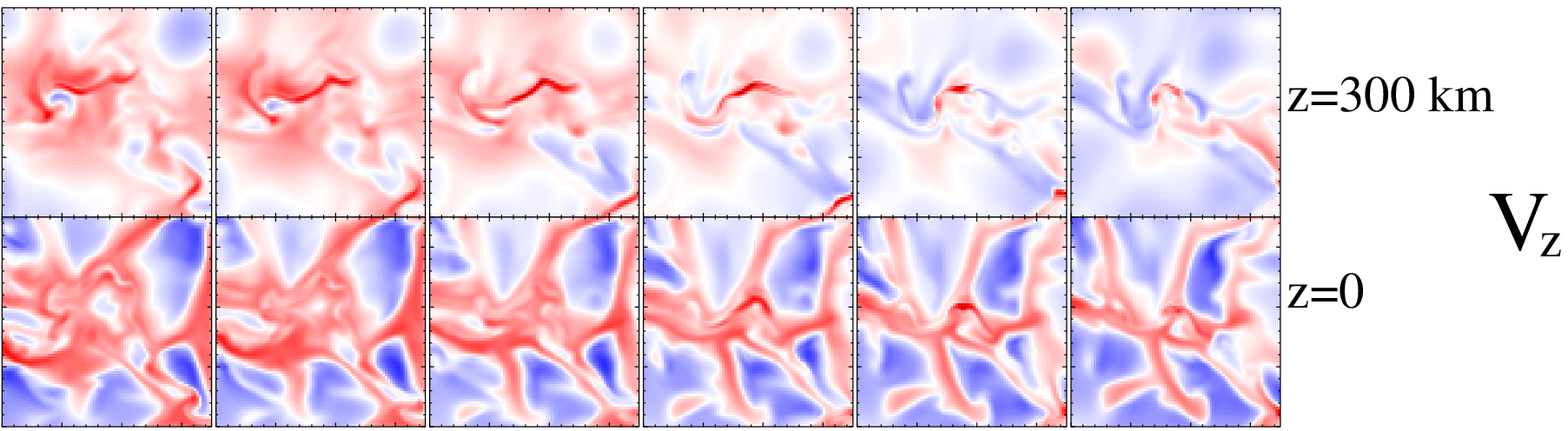}
\includegraphics[width=\hsize]{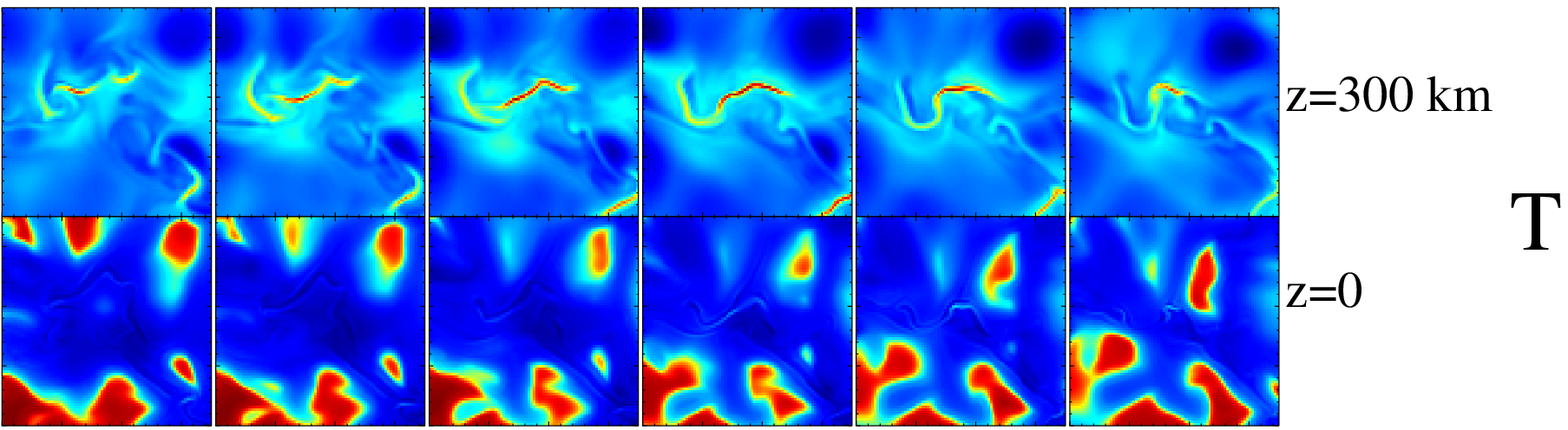}
\includegraphics[width=\hsize]{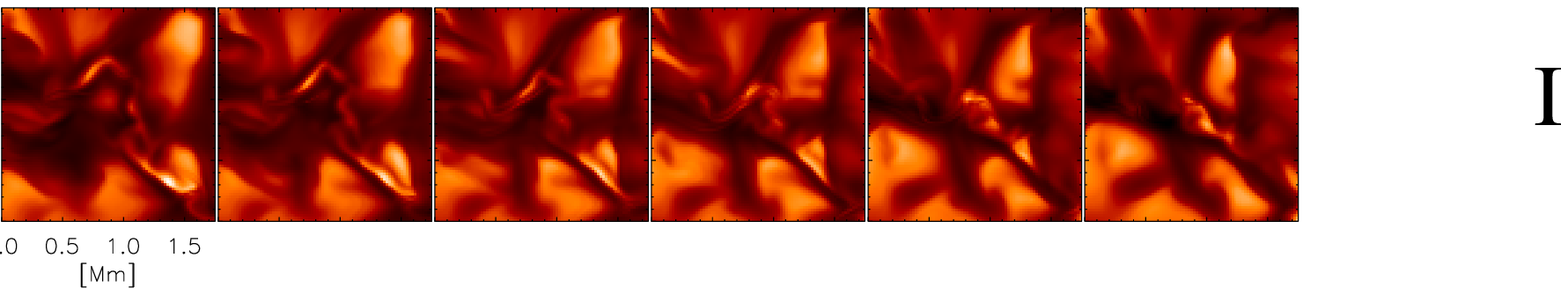}
\caption{Time series of a reconnection event (2x2 run, non-grey), 
approximately 30 minutes after introduction of magnetic field. 
The time between snapshots is 30 seconds. From top to bottom are
the vertical component of the magnetic field, the vertical component
of the velocity, the temperature and lastly the bolometric intensity. 
The first three quantities are shown at two heights, near $z=0$~km (lower
rows) and near $z=300$~km (upper rows). The horizontal extent of the maps 
is 1.75 $\times$ 1.75 Mm$^2$.  The intensity timeseries corresponds to 
the continuum intensity at 500 nm. The greyscale for the magnetic field 
covers the range from -530~G to +530~G at $z=300$~km and from -1760~G to
1760~G at z=0; the velocity range is from -7~km/s to 7~km/s at $z=300$~km
and -9~km/s to 9~km/s at $z=0$; the temperature range is from 4000$^{\circ}$ (blue)
to $7500^{\circ}$ (red) at $z=300$~km and 5000$^{\circ}$ to 10000$^{\circ}$ at $z=0$;
and the intensity varies between 0.65 and 1.5 of the averge quiet Sun value. }
\label{fig:event}
\end{figure*}

An example of such a reconnection event is shown in Figure~\ref{fig:event} ,
where two opposite-polarity magnetic features have been brought into contact
in an intergranular lane.

\begin{figure}[h!]
 \centering
\resizebox{1.0\hsize}{!}{ 
\includegraphics{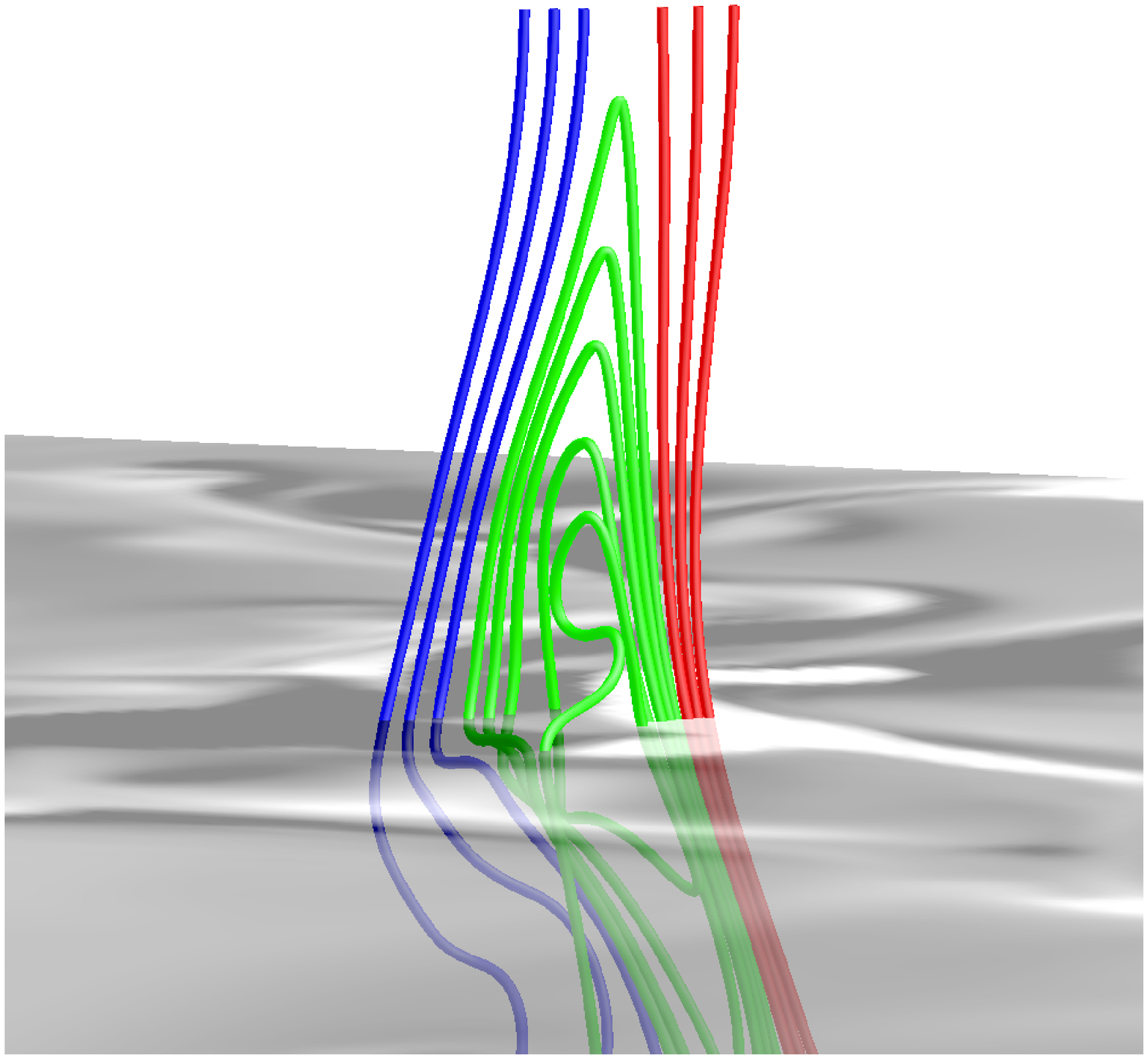}
\includegraphics{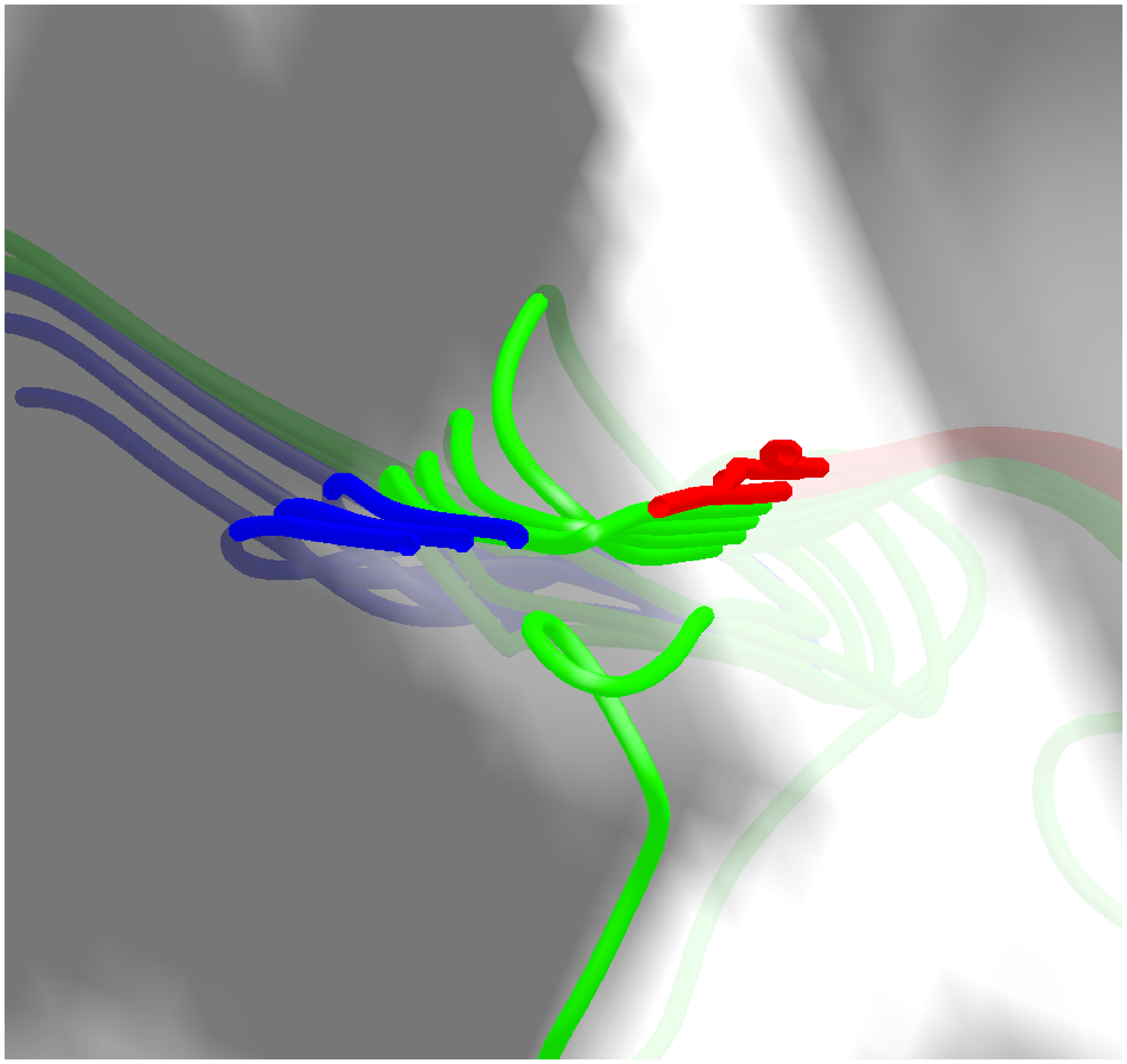}}
\caption{The colored lines are a 3D visualization of field lines at beginning 
of the reconnection event shown in the \ref{fig:event}. The grey-scale 
corresponds to the vertical component of the field at $z=0$. 
The left panel is a side view, the right panel is a top view.}
\label{fig:arc}
\end{figure}

The reconnection, which occurs above the surface,
has locally heated the plasma. The strong down flows in the upper photosphere
are the result of the retraction of {\iuloop}s (see Figure~\ref{fig:arc}) 
due to magnetic tension. Since the reconnection occurs close to the upper boundary,
the corresponding {\uloop}s are rapidly removed from the system so that their signature 
in this reconnection event is very weak.
 

A detailed description of the process involved and their observational signatures 
has been given by \cite{Danilovic09}. 
The time scale for the individual reconnection events is of the order of 10 minutes, which 
is short compared to the decay rates found here. In what follows we concentrate on the 
statistical properties of the decay of the field rather than on the detailed reconnection 
processes. To do so we consider magnetic field lines, 
which are described by the function ${\bf X}$, with  
\begin{equation}
\frac{d {\bf X}(s,x_0,y_0,z_0)}{d s}= \frac{{\bf B}\left[ {\bf X}\left(s\right)\right]}
                                  {\mid{\bf B}\left[ {\bf X}\left(s\right)\right] \mid + \epsilon} 
\end{equation}
and
\begin{equation}
{\bf X}(s,x_0,y_0,z_0)= (x_0,y,0,z_0).
\end{equation}
where $\epsilon=10^{-5}$ prevents problems where ${\bf B}=0$ and 

$s$ is a coordinate along the field line, and $x_0,y_0,z_0$ is point in space corresponding to $s=0$
and the constant $\epsilon=10^{-5}$~G prevents problems where ${\bf B}=0$.
In regions where $B \gg \epsilon$, $s$ is the length along the field line.

For our analysis we consider the set of field lines passing through the centre of each 
grid cell, i.e., we consider $288 \times 288 \times 100$ values of
$\{x_0,y_0,z_0\}$. This mean that near $t=0$ each cell will be sampled by 100 field lines,
each of which will pass through the centre of at least one cell.
As time progresses stronger field regions will become even more oversampled and
weak-field regions less sampled. At all times however each cell will be sampled by at least
one field line. Through this we are able to capture
cells which contain field lines entering and leaving the computational box 
through different boundaries.  
For each field line, we integrate Eq. 3 numerically 
using a fixed step size $\Delta s=3.5$~km until the field line exits
the domain through either the upper or lower boundary\footnote{The integration was limited 
to the range $-3500km <s< 3500$~km. Since the computational domain is only $1.4$~Mm in height most field lines
left the box within this range.}. 
Since the box is periodic in the
horizontal directions almost all
of the field lines pass through the upper or lower boundary.

The connectivity of the field line is defined by the boundary where it enters and leaves the 
computational box. A connectivity map is then created by considering
the connectivity of all field lines.
Since initially the field lines are all purely vertical, they can be classified as passing from the
top to the bottom of the box or from the bottom to the top. 
As reconnection take place, some of these field lines are converted into 
{\uu} and {\iuloop}s. Cells can contain field lines with different types of
connectivities.

\begin{figure}[h!]
\centering
\resizebox{0.8\hsize}{!}{ \includegraphics{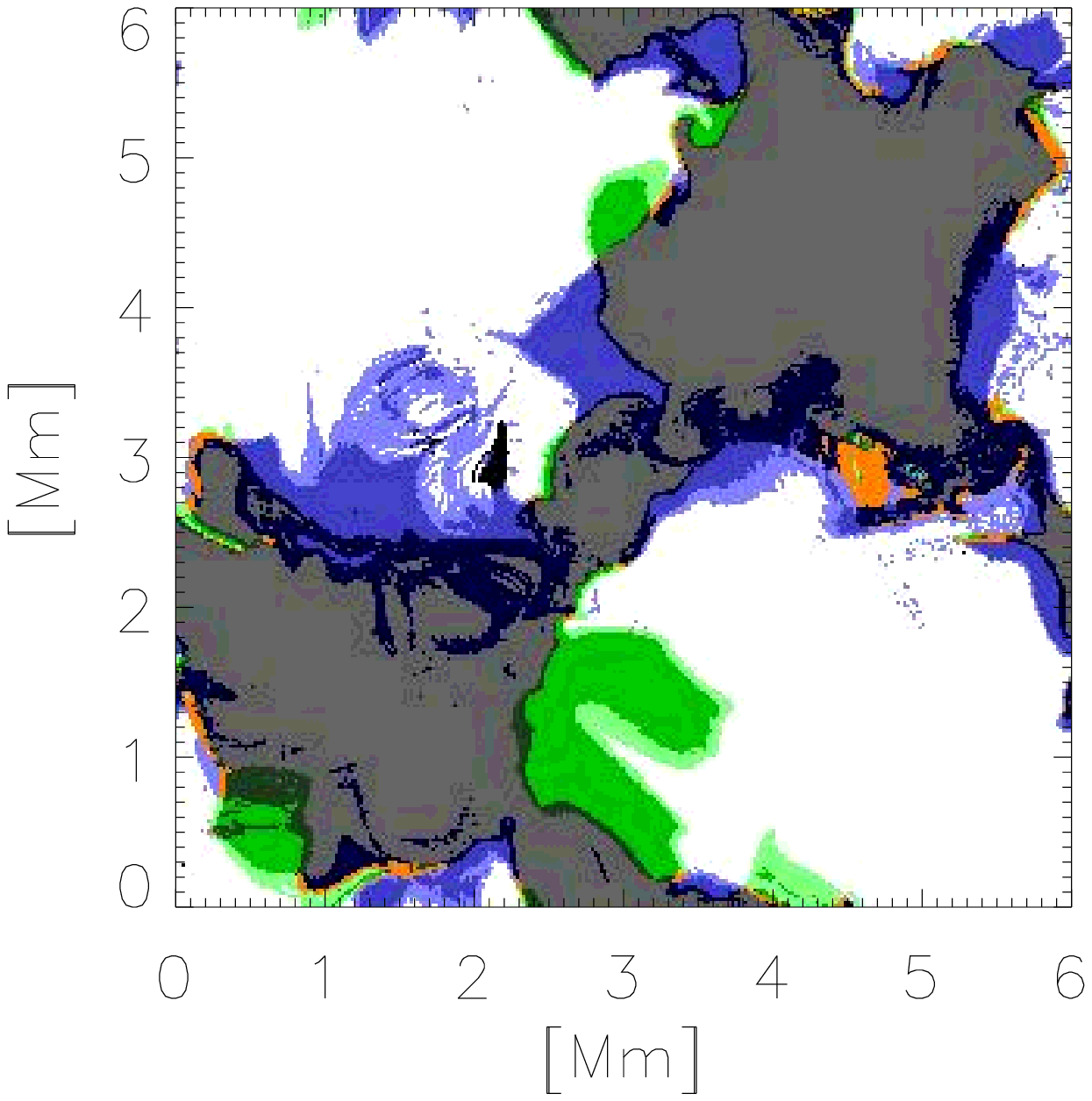}\includegraphics{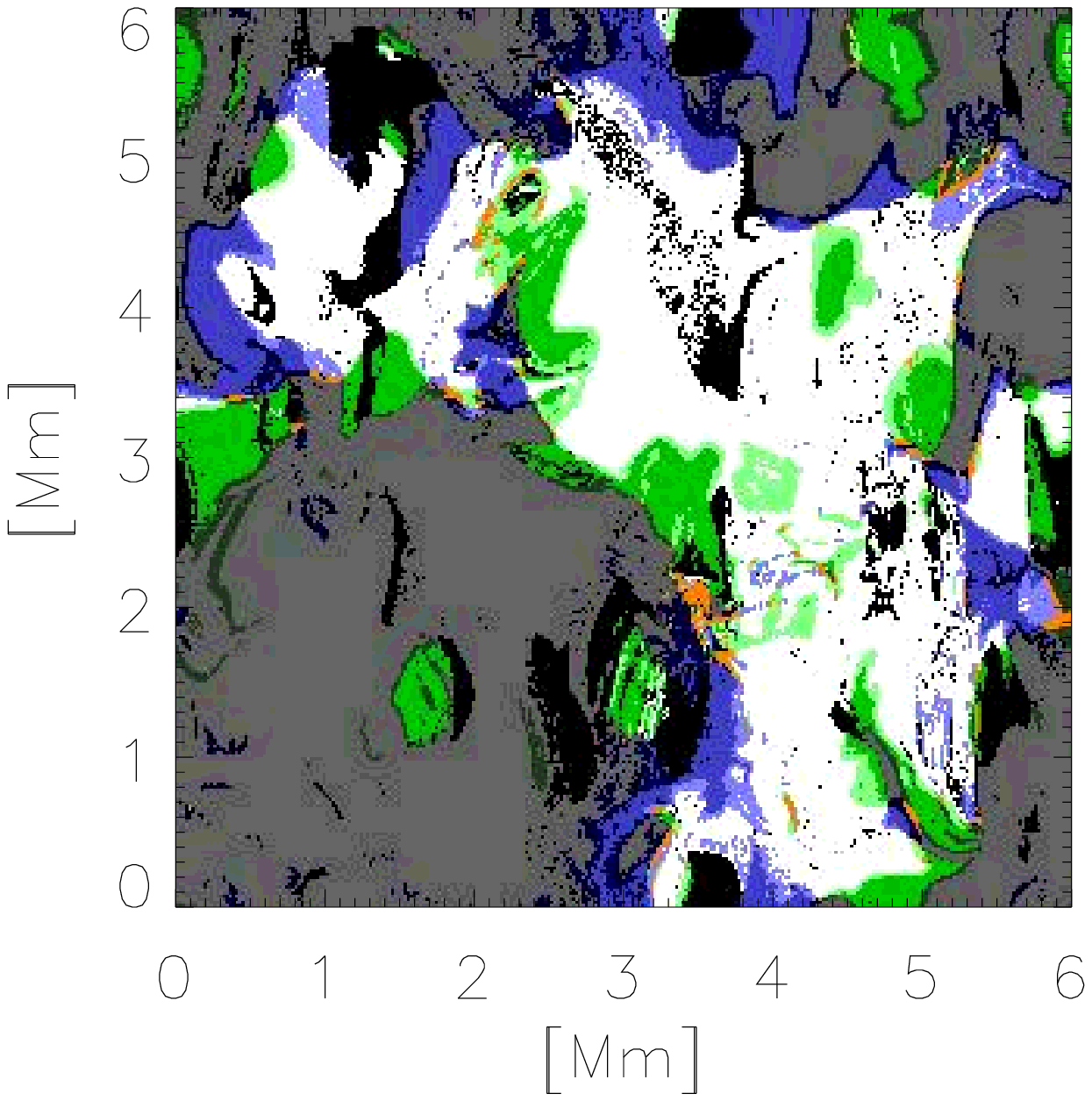}}
\resizebox{0.8\hsize}{!}{ \includegraphics{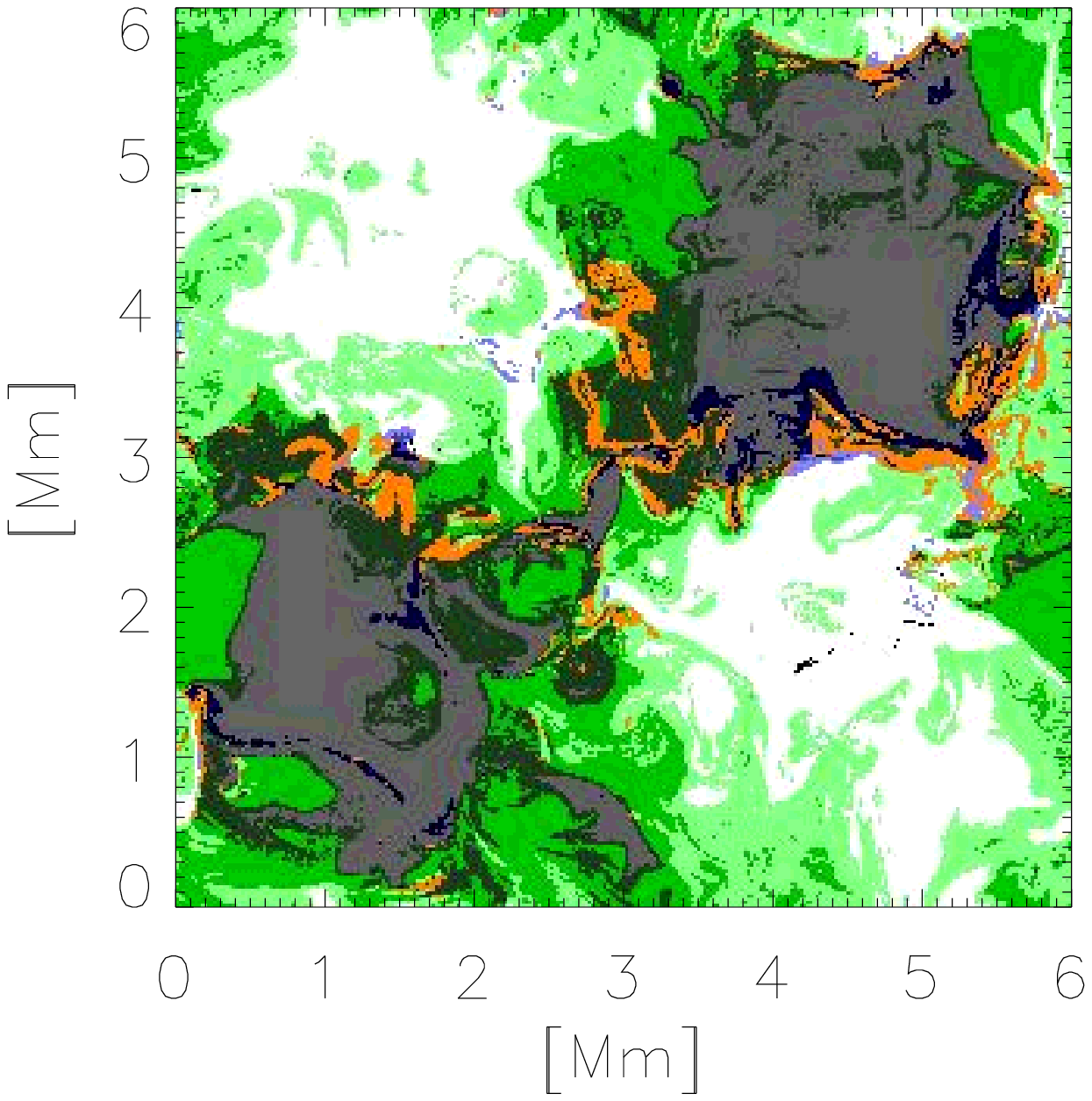}\includegraphics{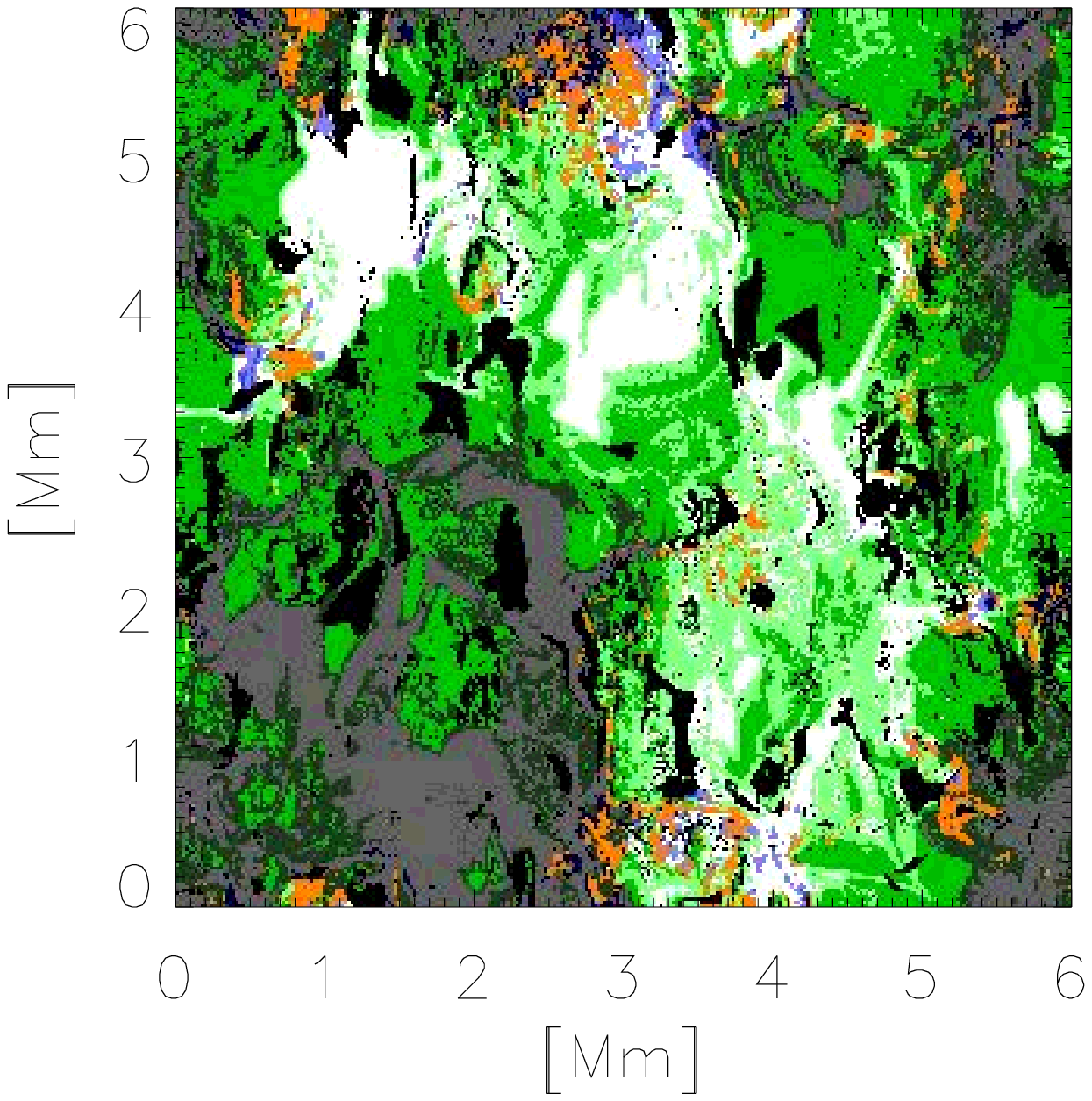}}
\resizebox{0.8\hsize}{!}{ \includegraphics{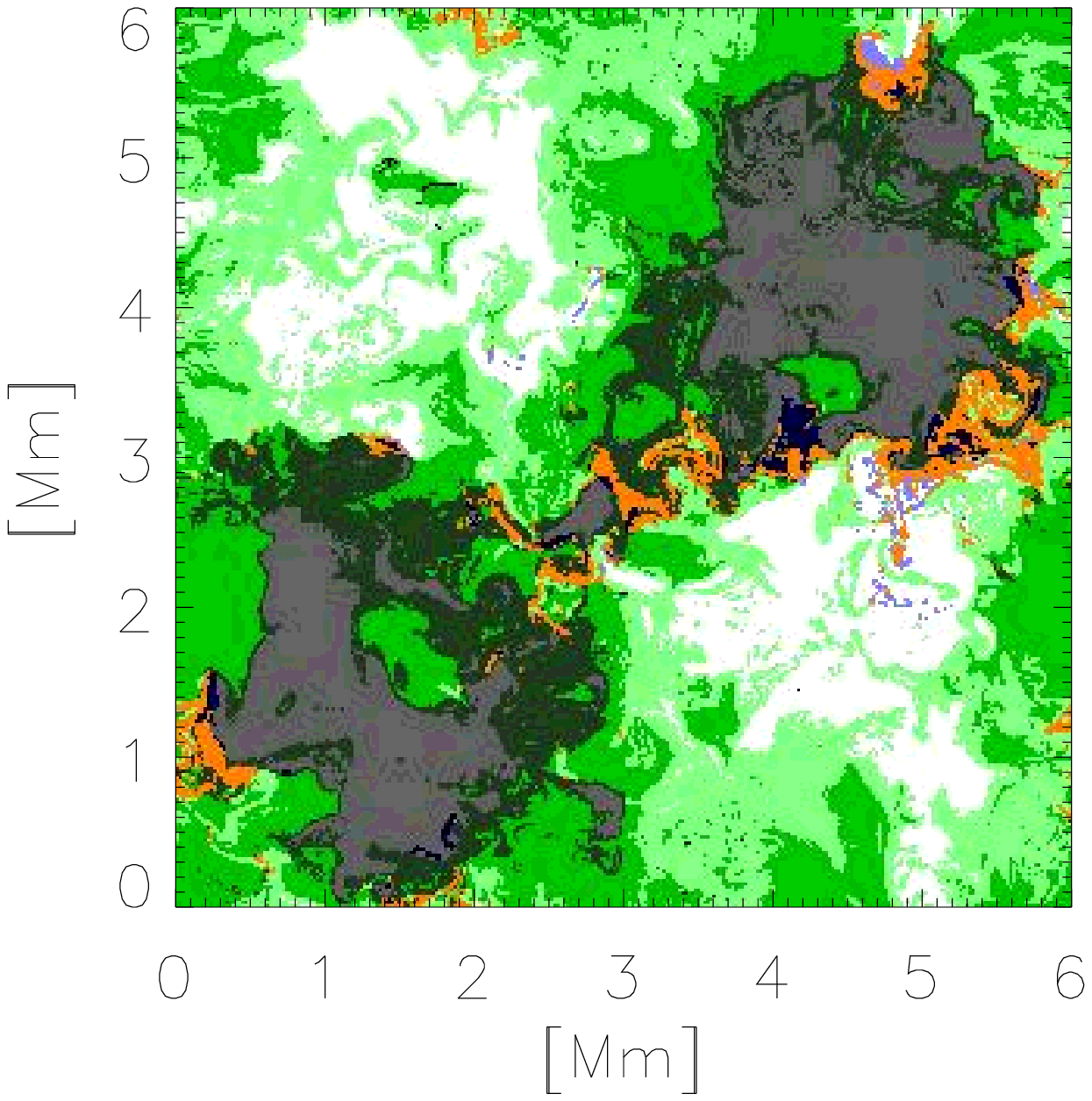}\includegraphics{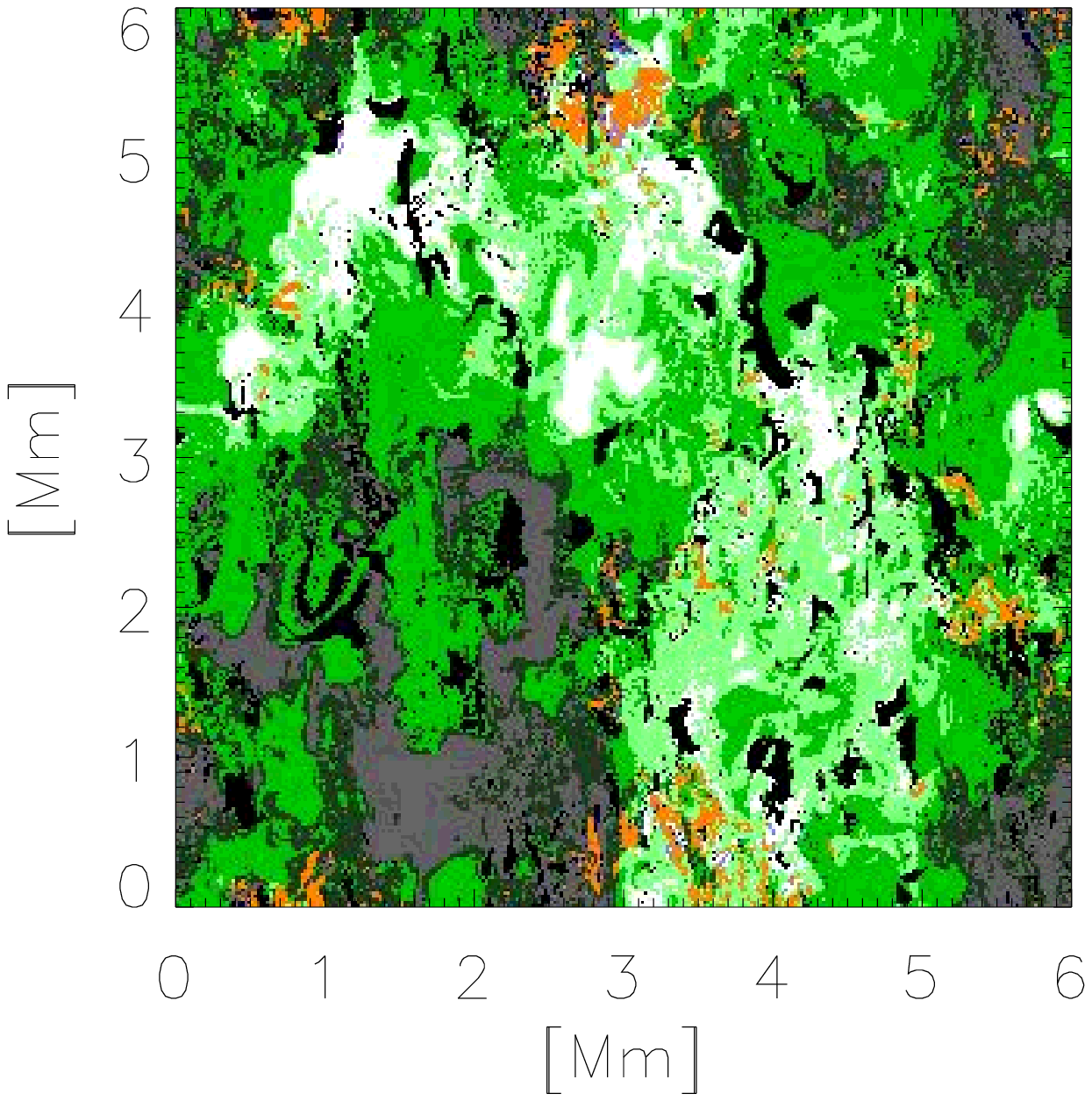}}
\caption{Field line connectivities at 3 heights (from bottom to top:
$z=-420$~km, $z=0$ and $z=420$~km). As described in the text, field lines were traced from
the centre of each of the $288 \times 288 \times 100$ grid cells, so that on average each 
cell contains many field lines.
The left panels correspond to  $t=16$ minutes, the right panels to $t=142$ minutes
of simulated time.
White indicates areas where field lines pass from the bottom of the
box to the top. Grey shows areas where field lines run from the top of the box to the 
bottom. Green indicates cells where all field lines have both legs passing through the 
bottom of the box ({\iuloop}s).
Blue indicates {\uloop}s with both legs passing through the upper surface.
Light (dark) green indicates the presence of both {\iuloop}s and field lines running from the
bottom of the box to the top (top to bottom), within one grid cell. Light and dark blue regions are
similarly defined but for {\uloop}s. The orange regions indicate where at least three types of field lines 
pass through a single grid cell.}
\label{fig:topology-cuts}
\end{figure}

The loops can then retract and leave through the boundaries 
of the simulation domain.  Figure~\ref{fig:topology-cuts} illustrates the connectivity of
the field lines at three different heights for a time just after the initial period of 
flux expulsion and late in the simulation (at 16 and 142 minutes, respectively).  
The plane at height $z=420$~km is in the upper photosphere,
where the density has fallen considerably so that the magnetic tubes fan out and become 
smooth and space filling. The lowest layer, at  $z=-420$~km, is within the convection 
zone, where the plasma motions dominate the magnetic field. Hence the magnetic field is more
tangled and the connectivity map is more complicated.

In the connectivity maps 16 minutes, the initial checkerboard arrangement is 
still visible at all heights. Inverse-$U$ loops, both ends of
which connect through the lower boundary are present mainly near the
borders of the opposing polarities, indicating that reconnection has taken place.
The {\uloop}s, both ends of which leave through the upper boundary 
mainly occur in the layers above the visible surface. 
The maps at 142 minutes
are similar, except that the {\iuloop}s are present over an even wider region of space and
the initial checkerboard pattern is less obvious.

The predominance of {\iuloop}s over {\uloop}s is a consequence of the fact that 
the reconnection occurs mainly above the optical surface,
where the plasma beta is low and the magnetic field concentrations, expanding 
to become space filling, first come into contact with each other.
The {\uloop}s then contain
little mass and are quickly pulled upwards by the comparitatively 
strong tension force. The {\iuloop}s, on the other hand, once they are 
pulled below the $\tau=1$ level, are loaded with the mass
from the subphotosphere and the tension force does not dominate over the other forces
in this subsurface region. Magnetic tension is dominant in the current sheets in 
the low-$\beta$ region above the surface.

\begin{figure}[h!]
\resizebox{0.8\hsize}{!}{\includegraphics{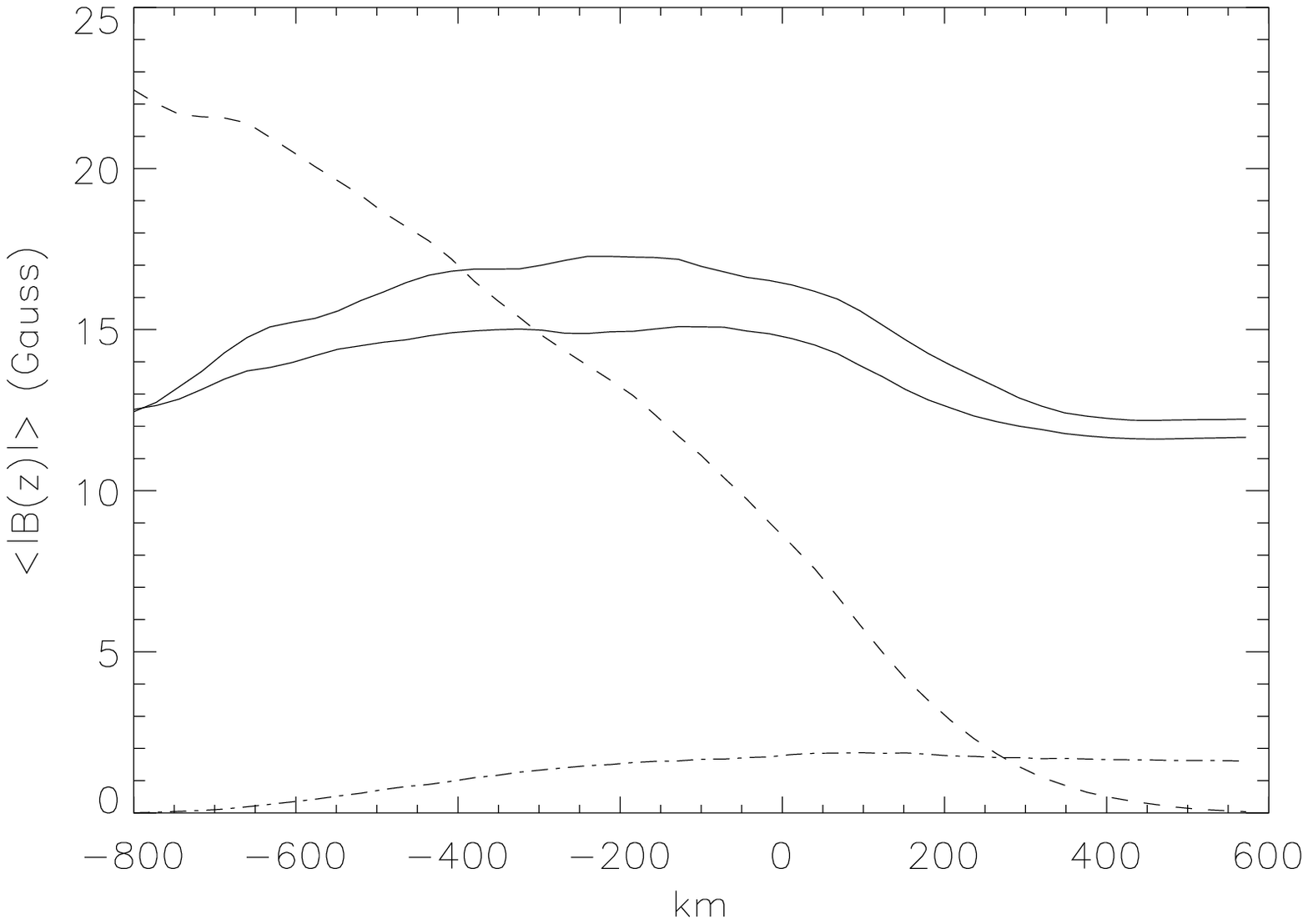}}
\resizebox{0.8\hsize}{!}{\includegraphics{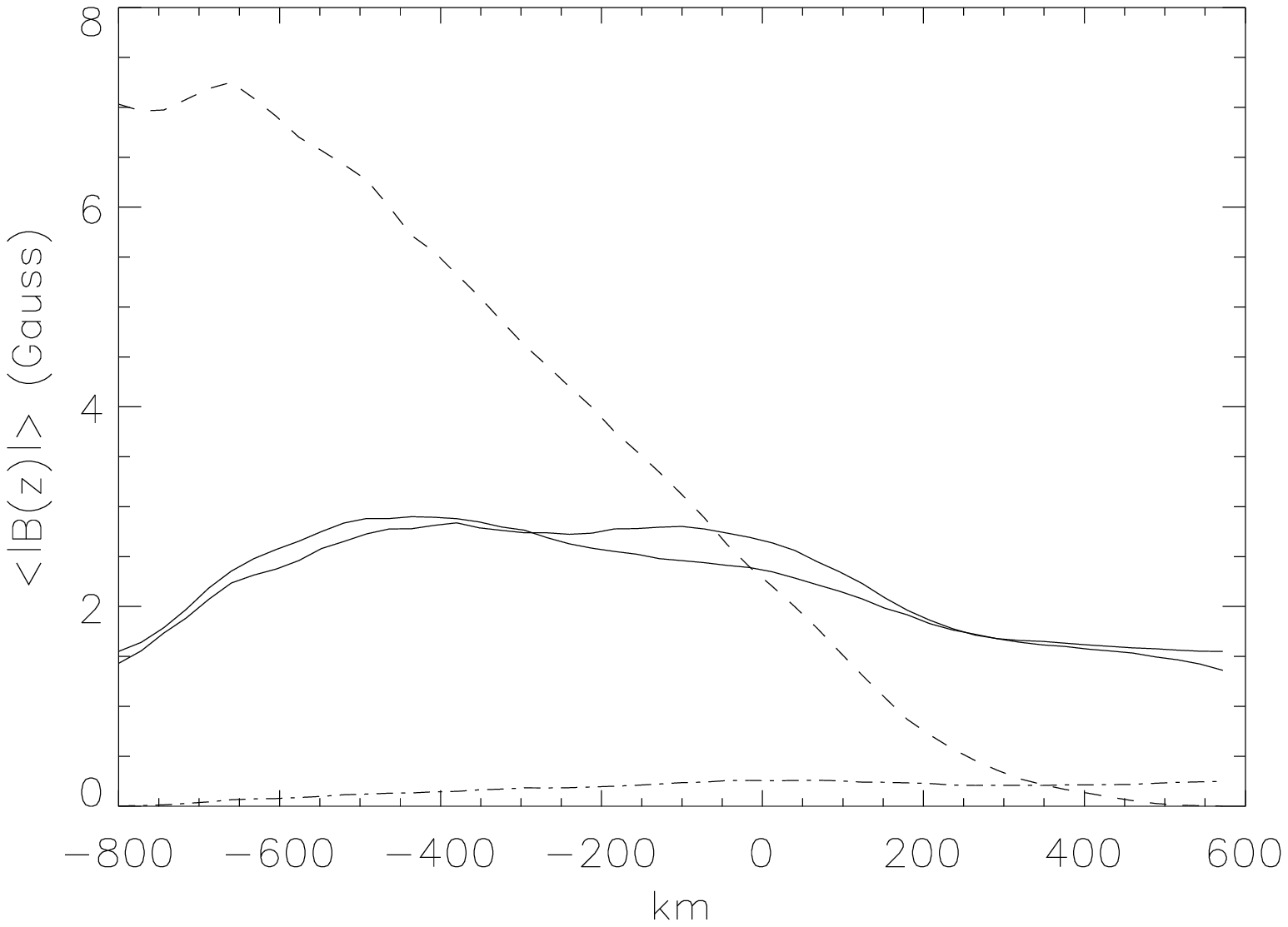}} 
\caption{Unsigned flux density as a function of height for the four
connectivities.  The solid lines correspond to the top-bottom and bottom-top connectivities, the dashed lines  
to {\iuloop}s, and the dot-dashed line to {\uloop}s. 
The upper panel is for $t=16$ minutes, soon after the initial transient phase. 
Already a substantial amount of {\iuloop} flux has been generated. The lower panel
is for $t=142$ minutes. The top-bottom and bottom-top connectivities use the same linestyle
to reflect the $B\rightarrow (-B)$ symmetry of the MHD equations.  
}
\label{fig:topology-profile}
\end{figure}

The retraction of the {\iuloop}s is thus a much slower process. This explains
the time evolution of the amount of flux in the different types of connectivities shown 
in Figure~\ref{fig:topology-profile}. Between $t=16$~min and $t=142$~min,  the amount of 
unsigned flux corresponding to each type of connectivity at all heights has fallen substantially.
The slowest decay is seen for the {\iuloop}s, where the unsigned flux has fallen by a 
factor of approximately 3 below the $\tau=1$ height and by a factor of 3-5 above.  
The unsigned flux associated with top-to-bottom and bottom-to-top 
field lines in the same time has fallen by a factor of approximately 7, and the decrease in the flux
of the {\uloop}s is even more dramatic. This reflects the fact that removal of {\uloop}s is a much
faster process than the removal of {\iuloop}s.
Importantly, if the 
reconnection were mainly occuring in the sub-photospheric region, then the
{\uloop}s would also be loaded with substantial mass and would only escape much more 
slowly, so that they would be more abundant.

\subsection{2$\times$1 initial condition.}

\begin{figure}[h!]
\resizebox{1.0\hsize}{!}{ 
\includegraphics{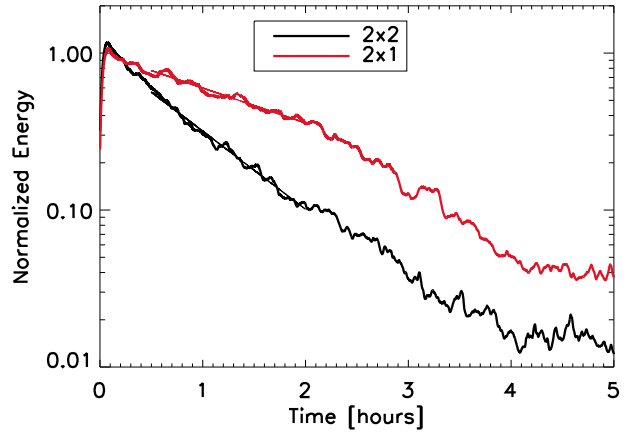}}
\caption{Time evolution of the total magnetic energy starting from the 2$\times 1$ (striped) 
initial condition. The curve from the 2$\times 2$ (reference) case is reproduced for comparison.}
\label{fig:Time_evol_2x1}
\end{figure}

We now consider the value of $\eta_{\mathrm{eff}}$ based on the evolution from an initial
condition consisting of two stripes ($2\times 1$) of opposite-polarity magnetic field.  
Again we quantify the decay rate by considering the decay of the total magnetic
energy in the computational domain. This is plotted in Figure~\ref{fig:Time_evol_2x1}. The $e$-folding
time for the decay is found to be 1hr 57 min. The dominant wavelength is $6000$~km,
so that the estimate for the turbulent diffusivity becomes $\eta_{\mathrm{eff}}= 129$~km$^2$s$^{-1}$. 


\subsection{Upper boundary conditions}
We also consider the effect of varying the upper boundary condition on the decay rate. For the magnetic field we assume  vertical
or matched-to-potential boundary conditions; for the flows we consider either closed
or open conditions, following \cite{Stein98} and \cite{Bercik02}. We also vary the height at which the boundary condition is imposed
($z \approx 600$~km above the $\tau=1$ level to $z \approx 840$~km). In each case, we use the 
$2\times 2$ initial condition as in the reference case.   

\begin{figure}[h!]
\resizebox{1.0\hsize}{!}{ 
\includegraphics{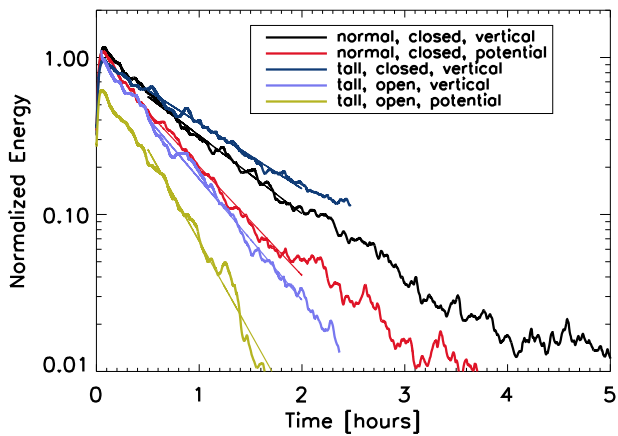}}
\caption{Time evolution of the total magnetic energy for the 2$\times 2$ case with
several different upper boundary conditions located at one of two heights.
The `normal, closed, vertical' case refers to our standard 2x2 run. The
`normal, closed, potential' refers to a case where the magnetic field was 
matched to a potential field at the top boundary. For the `tall' cases the 
top boundary was moved upwards by $280$~km, and `open' refers to cases where 
flow, temperature and pressure were smoothly extrapolated through the 
upper boundary.}
\label{fig:Time_evol_pot}
\end{figure}

The decay of the total magnetic energy for the different cases can be seen in
in Figure~\ref{fig:Time_evol_pot}.  The $e$-folding time for the magnetic 
energy and the estimated turbulent diffusivities are listed in Table~1. The 
values for the turbulent diffusivity resulting from the use of different upper boundary 
conditions vary by a factor of 3, from 145~km$^2$s$^{-1}$ for the vertical field, closed
boundary applied at $z\approx 600$~km to 341~km$^2$s$^{-1}$ for the potential field,
open boundary applied at $z\approx 840$~km. 
\begin{table*}
\caption{Summary of Magnetic energy decay rates and $\eta_{\text{eff}}$} 
\label{table:1}      
\centering            
\begin{tabular}{l l l l l r r}
\hline\hline 
    IC         &height &field BC&flow BC&$\eta$ (~km$^2$s$^{-1}$) &$e$-folding time (min) & $\eta_{\text{eff}}$ (km~$^2$s$^{-1}$)\\  
\hline 
    $2\times 2$ &normal&vertical &closed &11&  52 & 145\\
    $2\times 1$ &nomal &vertical &closed &11& 117 & 129\\
    $2\times 2$ &normal&potential&closed &11&  37 & 202\\
    $2\times 2$ &tall  &vertical &closed &11&  62 & 123\\
    $2\times 2$ &tall  &vertical &open   &11&  34 & 225\\
    $2\times 2$ &tall  &potential&open   &11&  22 & 341\\
    $2\times 2$ &normal&vertical &closed &5.6& 75 & 101\\
\hline                                   
\end{tabular}
\end{table*}


\subsection{Explicit magnetic diffusivity parameter, $\eta$}
\begin{figure}[h!]
\resizebox{1.0\hsize}{!}{ 
\includegraphics{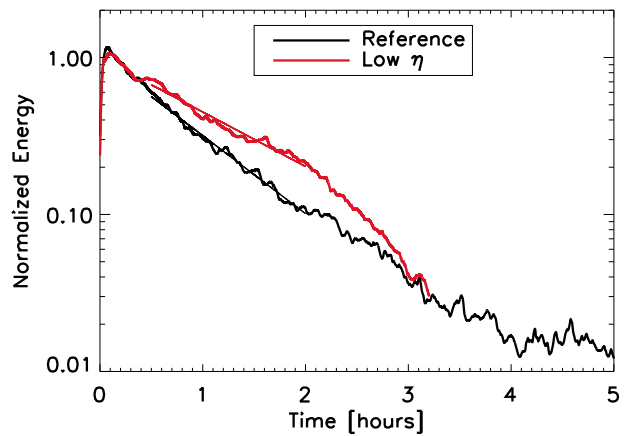}}
\caption{Time evolution of the total magnetic energy for a checkerboard (2$\times 2$) run
with $\eta=5.6$ as a function of time. The same curve from the reference case is shown for comparison.}
\label{fig:Time_evol_hir}
\end{figure}

We investigated the role of the explicit value of the input parameter $\eta$ by performing a 
simulation with twice as many grid points in the horizontal directions and with 
$\eta=5.6$~km$^2$s$^{-1}$.
The initial condition was the same as in the reference case, and the time evolution is
shown in Figure~\ref{fig:Time_evol_hir}. The fit to the total energy corresponds 
to an $e$-folding time of 1hr 15 min, and hence an effective diffusivity 
$\eta_{\mathrm{eff}}=101$~km$^2$s$^{-1}$. We note that the the fit to the curve in 
Figure~\ref{fig:Time_evol_hir} is sensitive to the time interval
over which the fit was made. Fitting over the interval 30--180 minutes leads to essentially the same 
estimate for $\eta_{\rm{eff}}$ as was obtained for the $2\times2$ reference case.
In this regard it is worth noting that the late-stage evolution in most cases
is different from that in the first 2~hours or so. This is presumably due to the fact that
the random motions move energy from those present in the initial condition modes (e.g. 2$\times$2) into other modes
(e.g. 1$\times$2 or 3$\times$2). When these modes dominate the decay, a different decay rate is to be expected.

\section{Discussion}
The values of $\eta_{\text{eff}}$ derived from the various runs and listed 
in Table 1 are within the range deduced from 
observations over a much larger spatial extent and longer time period 
\citep[see][Table 6.2]{Schrijver_book}.
To understand the differences between the estimated values of $\eta_{\mathrm{eff}}$
(i.e. the decay rates) for the different simulation setups,  
we note that the removal of flux in these simulations
is primarily determined by the rate at which opposite polarity patches are brought 
close enough to one another for the explicit diffusivity to become effective. The
differences between the values of $\eta_{\rm {eff}}$ can be understood in terms
of how close the magnetic polarities need to come for rapid diffusion to set in. In
loose terms this can be rephrased in terms of a `collisional cross section' of
the magnetic elements. When we use the potential boundary condition, the magnetic field
expands rapidly near the upper boundary, and hence magnetic elements which are still
separated near the surface can still come into contact in the upper layers. This is the 
reason why the reconnection mostly takes place above the surface.
It also explains,
qualitatively, why the potential boundary condition leads to a higher $\eta_{\rm eff}$
then was found in the reference case.

We may compare the values of $\eta_{\rm {eff}}$ with the estimate, 
$\eta_{\rm {turb}} \sim v_{\mathrm{rms}} L_{\mathrm{corr}}/3$, for the diffusion of
a passive scalar subject to random motions. In this estimate $v_{\mathrm{rms}}$ is the rms
velocity of the random motions and  $L_{\mathrm{corr}}$ is the correlation length scale
of the motions. 
The first question is then which motions to consider. Since we are mainly
concerned with the horizontal motions of the magnetic flux tubes we consider only the
horizontal velocity field in a $z=0$ slice from the simulations. The rms horizontal
velocity is then 2.4~km~s$^{-1}$. 
However, for the horizontal transport of the flux concentrations in the intergranular lanes,
the random horizontal motion of the granules is the relevant quantity. The 
corresponding rms value, for example as determined by  
local correlation tracking, is about a factor of two smaller than the
above value \citep{Rieutord01, Georgobiani07, Matloch10}, 
ie $v_{\mathrm{rms}}=1.2$~km~s$^{-1}$. 
For the correlation length of the horizontal velocity, we take $ L_{\mathrm{corr}}=1$~Mm, which is 
about the size of a granule. These values yield $\eta_{\rm {eff}}=400$~km$^2$~s$^{-1}$. 
This very simple, approximate value is slightly above the actual range of values found in the simulations.

\section{Conclusions}

We have studied the decay of simple arrangements of magnetic field in the near-surface
layers of the Sun. We found that
magnetic field is removed from the surface when magnetic elements of opposite
polarity, advected by the granular flows, come into close proximity above the surface. 
Reconnection then produces {\uloop}s, which quickly escape through the upper boundary, and 
 {\iuloop}s, which retract more slowly into the subphotosphere. 

During most of the decay phase, the relative contributions of magnetic energy coming from 
different field strengths are largely independent of the amount of flux. A weak field tail, 
however, does appear late in the evolution, when the unsigned flux levels are very low.

The decay of the magnetic field is reasonably well described by turbulent diffusion concept with
plausible values of $\eta_{\rm eff}$. The value of $\eta_{\mathrm{eff}}$  was found to depend on the upper 
boundary condition. The potential-field, open-boundary condition applied to the tall box is probably 
the most realistic in this regard.
This case has $\eta_{\rm eff}\approx 340$~km$^2$s$^{-1}$. The range of values from the different
simulations, from 100 to 340~km$^2$~s$^{-1}$, gives an indication of the uncertainty due
to, for example, the choice of the upper boundary condition. 

\bibliographystyle{aa}
\bibliography{pap_mixed}

\end{document}